\documentclass[debug]{rmaa}

\usepackage{paralist}

\usepackage{psfrag,color}

\usepackage[latin1]{inputenc}

\usepackage{aas_macros}
\usepackage{color}
\usepackage{array}
\usepackage{amsmath}
\usepackage{soul}
\usepackage{hyperref}
\usepackage{mathrsfs}

\title{On the correct computation of all Lyapunov exponents in Hamiltonian dynamical systems} 

\author{
  D. D. Carpintero,\altaffilmark{1,2} 
  and J. C. Muzzio\altaffilmark{1,2}}

\altaffiltext{1}{Facultad de Ciencias Astron\'omicas y Geof\'isicas, Universidad
Nacional de La Plata, Argentina.}

\altaffiltext{2}{Instituto de Astrof\'isica de La Plata, UNLP--Conicet,
Argentina.}

\shortauthor{Carpintero, \& Muzzio}
\shorttitle{Correct computation of Lyapunov exponents}

\fulladdresses{
\item Daniel D. Carpintero and Juan C. Muzzio: Facultad de Ciencias Astron\'omicas y Geof\'isicas, Universidad
Nacional de La Plata, Paseo del Bosque s/n, 1900 La Plata, Argentina (ddc, jcmuzzio@fcaglp.unlp.edu.ar).
\item Daniel D. Carpintero and Juan C. Muzzio: Instituto de Astrof\'isica de La Plata, UNLP--Conicet, Paseo del Bosque s/n, 1900
La Plata, Argentina (ddc, jcmuzzio@fcaglp.unlp.edu.ar).}

\listofauthors{D. D. Carpintero, \& J. C. Muzzio}
\indexauthor{Carpintero, D. D.}
\indexauthor{Muzzio, J. C.}

\abstract{The Lyapunov Characteristic Exponents are a useful indicator of chaos in astronomical dynamical systems.
They are usually computed through a standard, very efficient and neat algorithm published in 1980. However,
for Hamiltonian systems the expected result of pairs of opposite exponents is not always obtained with enough
precision. We find here why in these cases the initial order of the deviation vectors matters, and how to sort
them in order to obtain a correct result.}

\resumen{Los Exponentes Caracter\'isticos de Lyapunov constituyen un \'util indicador de
caos en sistemas
din\'amicos astron\'omicos. Habitualmente se los calcula mediante un algoritmo
muy claro y eficiente
publicado en 1980. Sin embargo, para sistemas hamiltonianos, el resultado
esperable de pares de
exponentes opuestos no siempre se obtiene con suficiente precisi\'on. Aqu\'i
encontramos
por qu\'e en estos casos es importante el orden inicial de los vectores de
desviaci\'on,
y c\'omo deben distribuirse
a fin de obtener un resultado correcto.}

\addkeyword{galaxies: kinematics and dynamics}
\addkeyword{planets and satellites: dynamical evolution and stability}
\addkeyword{methods: numerical}
\addkeyword{chaos}

\begin{document}
\maketitle

\section{Introduction}
\label{intro}

The importance of chaos in astronomical dynamical systems is generally recognized nowadays
and a complete summary of this subject can be found in the textbook of
\citet{C02}. The algorithms usually used to determine the regularity or chaoticity of a dynamical
system can be grouped into two categories: those based on the frequency analysis of the orbits
\citep[e.g.][]{BS82,L90,SN96,CA98,PL98} and the so-called variational indicators, based on the
behaviour of deviation vectors \citep[e.g.][]{VC94,CV96,FGL97,VCE99,CS00,SEE00,S01,LF01,FLFF02,
CGS03,SESF04,SBA07,MDCG11,DMCG12,DMCG12p,MDCG13,CMD14}. Among the methods of the last group,
the computation of the Lyapunov characteristic exponents (LCEs) \citep{BGS76,BGGS80a,BGGS80b}
stands out not only for being the oldest of them all but also for representing the very definition
of chaos. They have been used to investigate chaos in elliptical galaxies, among others, by
\citet{UP88}, \citet{MF96} and by ourselves (see \citet{CM16} and its references to our previous work).

In the second part of a seminal paper, \citet{BGGS80b} gave for the first time a thorough description of an algorithm to compute all the LCEs of a system, which in turn was based on the theoretical work of the first part \citep{BGGS80a}. This algorithm quickly became a standard.

However, when dealing with a Hamiltonian system, the expected result of paired opposite LCEs is not always achieved with enough numerical precision, notwithstanding the neat procedure of the above-mentioned algorithm. We find here the origin of the problem and an easy way out. 

\section{Setting the stage}

We assume that we are dealing with a dynamical system described by $n$ differential equations of the first order. To compute the LCEs of one of its orbits with the algorithm of \citet{BGGS80b}, one obtains the orbit itself ---which we will call the unperturbed orbit--- by integrating the corresponding equations of motion, plus the time evolution of $n$ linearly independent vectors, the deviation vectors $\delta\mathbf{w}_i$, $i=1,\dots,n$, representing the phase space distance between the orbit and $n$ additional orbits ---the perturbed orbits--- that start near the former. The time evolution of these vectors are obtained by integrating the so-called variational equations, that is, the first variation of the equations of the motion around the original orbit \citep[e.g.][p. 148]{T89}. To obtain the LCEs $\lambda_i$, $i=1,\dots,n$ from the deviation vectors, one computes \citep{BGGS80b}
\begin{equation}
\lambda_i \simeq \frac{1}{N \tau} \sum_{k=1}^N \ln
\alpha_i(t_k),
\label{calc}
\end{equation}
where $\alpha_i(t_k)$ is the modulus of $\delta\mathbf{w}_i$ after a orthogonalization of the set $\{\delta\mathbf{w}_j\}_{j=1,\dots,n}$ at time $t_k$, $\tau$ is the step of integration, and $N$ a positive integer which has to be large enough to get a good approximation. If the orthogonalization is carried out using the Gram-Schmidt method, the $\alpha_i$'s can be obtained on the fly. Also, since the computation does not depend on the initial moduli or initial orientation of the $\delta\mathbf{w}_i$ \citep{BGGS80a}, it is customary to set initial deviation vectors of unitary modulus, each one aligned with one of the Cartesian axis. 

Let us now assume, to fix ideas, that the LCEs are ordered in descending order according to their values:
\begin{equation}
\lambda_1 \ge \lambda_2 \ge \dots \ge \lambda_n.
\label{orden}
\end{equation}
Now we may ask: can we determine which of the $\delta\mathbf{w}_i$'s will give the $\alpha_i$ with which $\lambda_1$ is obtained? Clearly, by equation~(\ref{calc}), the deviation vector that accumulates the largest modulus after the orthogonalizations will be the one that gives $\lambda_1$. This vector is identifiable thanks to three circumstances. First, the orthogonalization at times $t_k$ includes a normalization of all the resulting vectors, in order to avoid numerical overflows due to their exponential growth. Thus, all the deviation vectors start their evolution always with the same modulus. Second, all the vectors tend to align towards the direction of maximum growth ---the reason why an orthogonalization is done periodically, thus avoiding very small angles between vectors which would be numerically intractable. Therefore, all vectors tend to go into a direction with the same rate of growth; this reason together with the previous one allow to claim that if the interval between $t_k$ and $t_{k+1}$ is not too large, all vectors will have similar moduli just before the orthonormalization step. Third, by using the Gram-Schmidt method, the first vector entering the algorithm will keep its modulus; the second one, instead, will end up with a smaller modulus because it is  projected into the subspace orthogonal to the first. The third, fourth, etc. vectors will end up with even smaller moduli, each one being projected into subspaces of lesser dimension. These three reasons together allow to answer the question posed above: although at short times any vector could be the largest, given enough time the first vector entering the Gram-Schmidt algorithm will grow more than the rest, and will be the one with which $\lambda_1$ will be computed. 

By reasoning in the same way, one could claim that the second vector entering the Gram-Schmidt algorithm will always have the second-largest modulus and therefore will be the one with which $\lambda_2$ will be computed, and so on for the rest of deviation vectors and LCEs. Therefore, if we sort the subindices of $\delta\mathbf{w}_i$ in the order with which they enter the Gram-Schmidt routine (i.e., $\delta\mathbf{w}_1$ the first one, $\delta\mathbf{w}_2$ the second one, etc.), then each $\delta\mathbf{w}_i$ should give the corresponding $\lambda_i$ ---the latter sorted according to inequation~(\ref{orden}). Although this is to be expected, it turns out that is not quite true in all cases.

\section{The problem: Hamiltonian systems}

\begin{figure}
\includegraphics[width=0.90\textwidth]{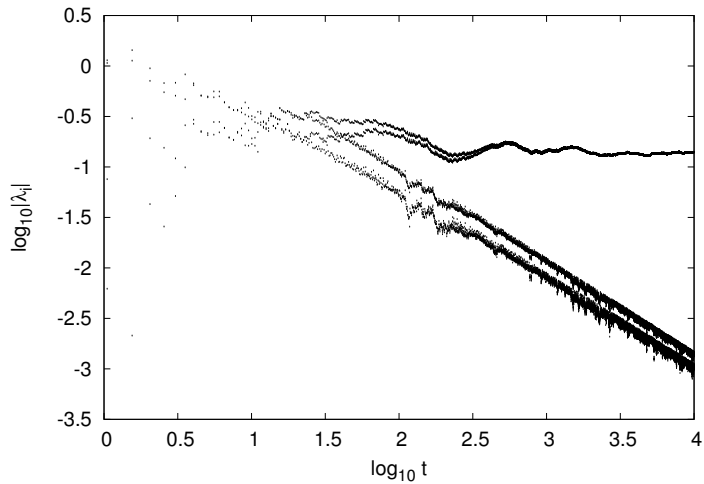}
\caption{LCEs of an orbit in the potential of equation~(\ref{pot}) with parameters $v_0=1$, $q=0.9$, $R_c=0.14$ and $R_e=3$, started at the phase space point $(x,y,\dot x,\dot y)=(0.1, 0.5, 0, 1)$. From top to bottom, the sets of dots are generated by deviation vectors originally pointing along the directions ${\bf e}_x$, ${\bf e}_y$, ${\bf e}_{\dot x}$ and ${\bf e}_{\dot y}$, respectively. All the computations were performed with double precision. Since we are plotting the absolute value of the LCEs, if two of them were the opposite of the other two ---as expected for a Hamiltonian system--- only two sets of dots would be seen.}
\label{mal}
\end{figure}

If the dynamical system under study is Hamiltonian and autonomous, by Liouville's theorem any volume of the phase space will be conserved along its evolution. From this is not hard to see \citep[e.g.][]{J89} that, for any direction of the phase space that stretches exponentially (with corresponding LCE positive), there must be another one that shrinks at the same exponential rate (with an LCE equal to the negative of the latter). Thus, Hamiltonian systems always have pairs of LCEs that are negatives of each other. Also, one of those pairs is always zero. These are strong restrictions that may be used to control whether the computation of the LCEs has been successful. But it turns out that a na\"ive application of equation~(\ref{calc}) does not always achieve this. Figure~\ref{mal} shows the absolute value of the computed LCEs of an orbit in the two-dimensional Binney potential \citep{B82}
\begin{equation}
\Phi(x,y) = \frac{v_0^2}{2} \ln\left[ R_c^2 + x^2 + \frac{y^2}{q^2} + \frac{1}{R_e}\sqrt{x^2+y^2} 
(x^2-y^2)\right],
\label{pot}
\end{equation}
where $(x,y)$ are Cartesian coordinates and $v_0$, $q$, $R_c$ and $R_e$ are parameters. Since the system is Hamiltonian and autonomous, we expect that $\lambda_1 = -\lambda_4$ and $\lambda_2 = -\lambda_3$, the latter tending to zero as time goes by. However, although the computation was done with double precision, we can see that the four LCEs are not well paired in opposite pairs.

\section{The solution}

An important property of Hamiltonian dynamics is that Poincar\'e invariants are preserved along the flow \citep[e.g.][]{T89}. Let \textbf{q} and \textbf{p} stand for the coordinates and momenta of a canonical set, respectively; in addition, let 
$\mathscr{C}_1$ be any closed curve in the $n$-dimensional phase space encompassing a tube of orbits, and $\mathscr{C}_2$ any other closed curve enclosing the same set of trajectories. The first Poincar\'e invariant can be expressed as
\begin{equation}
\int_{\mathscr{C}_1}  \mathbf{p}\cdot
\mathrm{d}\mathbf{q} =
\int_{\mathscr{C}_2}  \mathbf{p}\cdot
\mathrm{d}\mathbf{q}.
\end{equation}
We note that $\mathscr{C}_2$ could be the curve $\mathscr{C}_1$ evolved in time. Let us take, to fix ideas, $n=2$ and Cartesian coordinates $(\mathbf{q},\mathbf{p})=(x,y,\dot x,\dot y)$. If we choose for $\mathscr{C}_1$ the unit square in the plane $(q_1,p_1)=(x,\dot x)$ at $t_0$, it will enclose the area spanned by $\delta\mathbf{w}_x(t_0)$ and $\delta\mathbf{w}_{\dot x}(t_0)$, and the Poincar\'e first invariant will have only the term corresponding to $\dot x\cdot\mathrm{d}x$. Furthermore, by virtue of Stokes's theorem, the line integral will be equal to the enclosed area. If we choose the initial deviation vectors as before (unit vectors pointing along each Cartesian axis), this area will be equal to one, and by the invariance of the integral the area of the parallelogram spanned by $\delta\mathbf{w}_x(t)$ and $\delta\mathbf{w}_{\dot x}(t)$ for any $t>t_0$ will remain unitary.

Now, the orthogonalization of the Gram-Schmidt algorithm allows us to compute the area of the parallelogram spanned by $\delta\mathbf{w}_x(t)$ and $\delta\mathbf{w}_{\dot x}(t)$ as simply $\alpha_x \cdot \alpha_{\dot x}$, and since this area shoud be always equal to 1, we should have
\begin{equation}
\ln \alpha_x(t) = - \ln \alpha_{\dot x}(t)
\end{equation}
for all $t$. The same will happen with the $y$ coordinate. Looking at equation~(\ref{calc}), we see that the deviation vectors started on the $x$ and $\dot x$ axes will give a pair of opposite LCEs, while those started on the $y$ and $\dot y$ axes will give the other pair.

But here we see the problem. On the one hand, as we have just seen, if for example $\delta\mathbf{w}_1 = \delta\mathbf{w}_x$, then we must have $\delta\mathbf{w}_4 = \delta\mathbf{w}_{\dot x}$ in order to obtain the pair of opposite LCEs. But, on the other hand, this is not necessarily the order in which they are entered into the Gram-Schmidt routine. If these vectors are not the first and fourth, we are forcing the algorithm to find an opposite value of $\lambda_1$ with a vector that is not the expected one, thus compelling the routine to give the correct modulus to a vector that was not the one that formed a unitary parallelogram with the first. This is numerically inefficient.
Therefore, since the first and last deviation vectors inserted into the Gram-Schmidt routine should give the maximum LCE and its opposite sibling, they should be those originally pointing along a Cartesian coordinate and its corresponding velocity (in any order among them). The same for the second and penultimate vectors, etc. 

\begin{figure}
\includegraphics[width=0.90\textwidth]{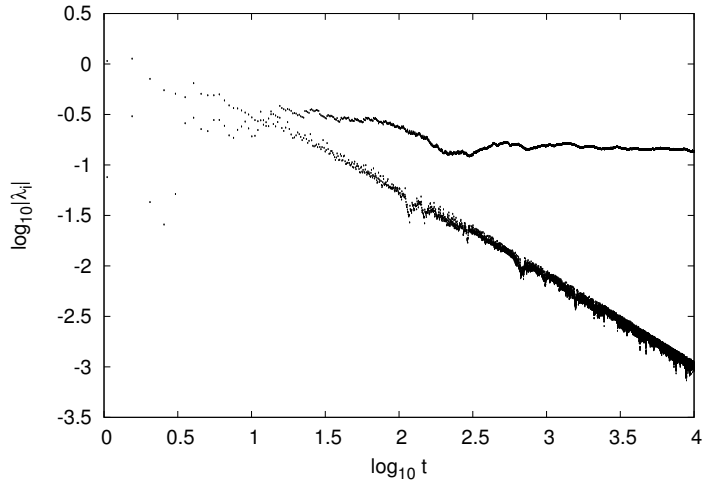}
\caption{LCEs of the same orbit of Figure~\ref{mal}, but starting with unitary deviation vectors $\delta\mathbf{w}_1=\mathbf{e}_x$, $\delta\mathbf{w}_2=\mathbf{e}_y$, $\delta\mathbf{w}_3=\mathbf{e}_{\dot y}$ and $\delta\mathbf{w}_4=\mathbf{e}_{\dot x}$. The top set of points were generated by $\mathbf{e}_x$ and $\mathbf{e}_{\dot x}$, though they are the same so a unique set is seen. The same for the bottom set, which is a superposition of the sets generated by $\mathbf{e}_y$ and $\mathbf{e}_{\dot y}$.}
\label{bien}
\end{figure}

Figure~\ref{bien} shows the LCEs of the same orbit as in Figure~\ref{mal}, but computed with the deviation vectors ordered originally in the directions $(\mathbf{e}_x,\mathbf{e}_y,\mathbf{e}_{\dot y},\mathbf{e}_{\dot x})$. There are four sets of points plotted in the figure, generated by the evolution of those vectors. But the sets generated by $\mathbf{e}_x$ and $\mathbf{e}_{\dot x}$ are superimposed, so only one set is visible (the top one in the figure). In the same way, the two sets generated by $\mathbf{e}_y$ and $\mathbf{e}_{\dot y}$ are superimposed (the bottom set in the figure). Thus, we can see that the very same algorithm with the same orbit used before gives now exact opposite LCEs, as expected.

\begin{figure}
\includegraphics[width=0.45\textwidth]{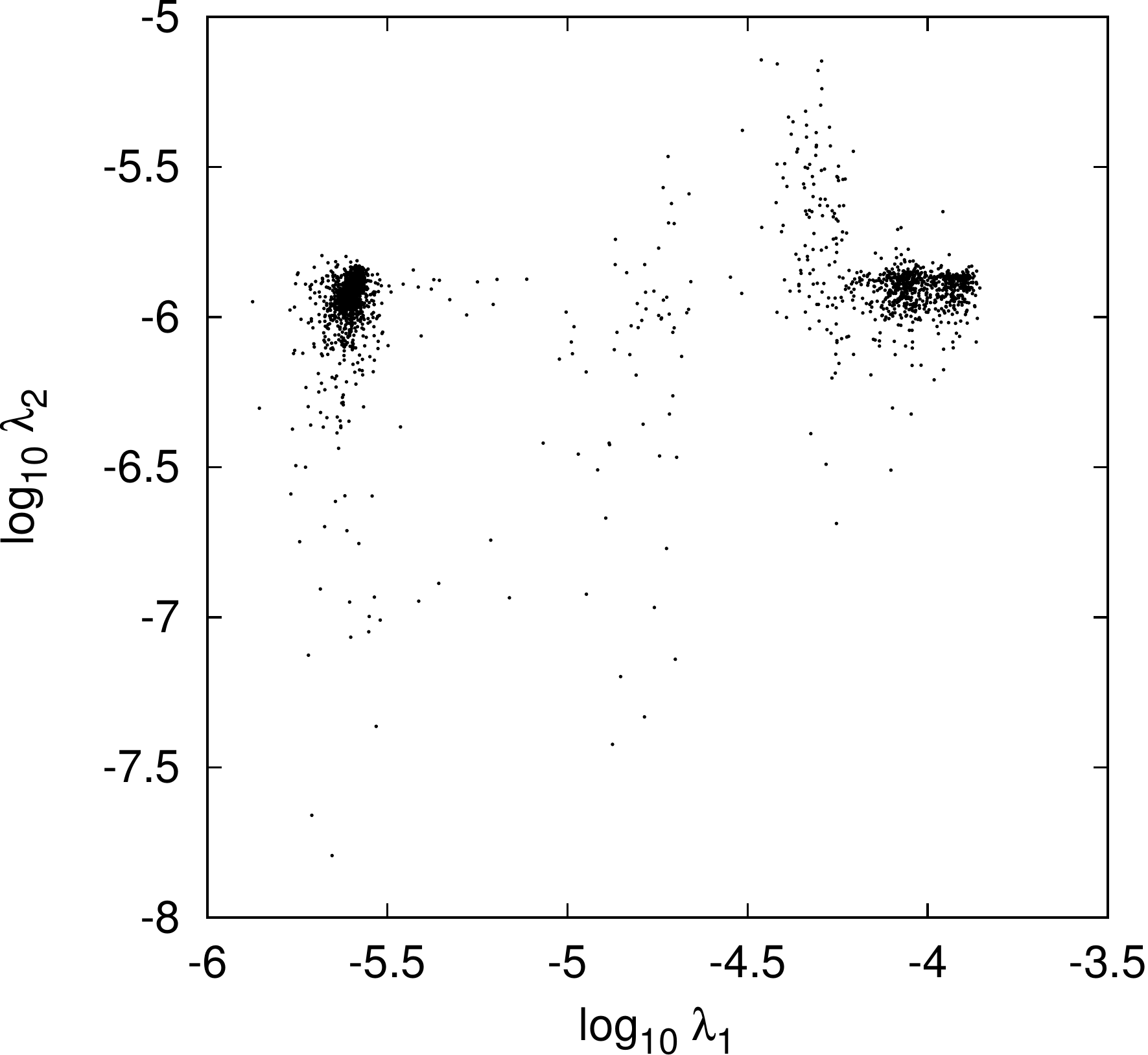}%
\hfill
\includegraphics[width=0.45\textwidth]{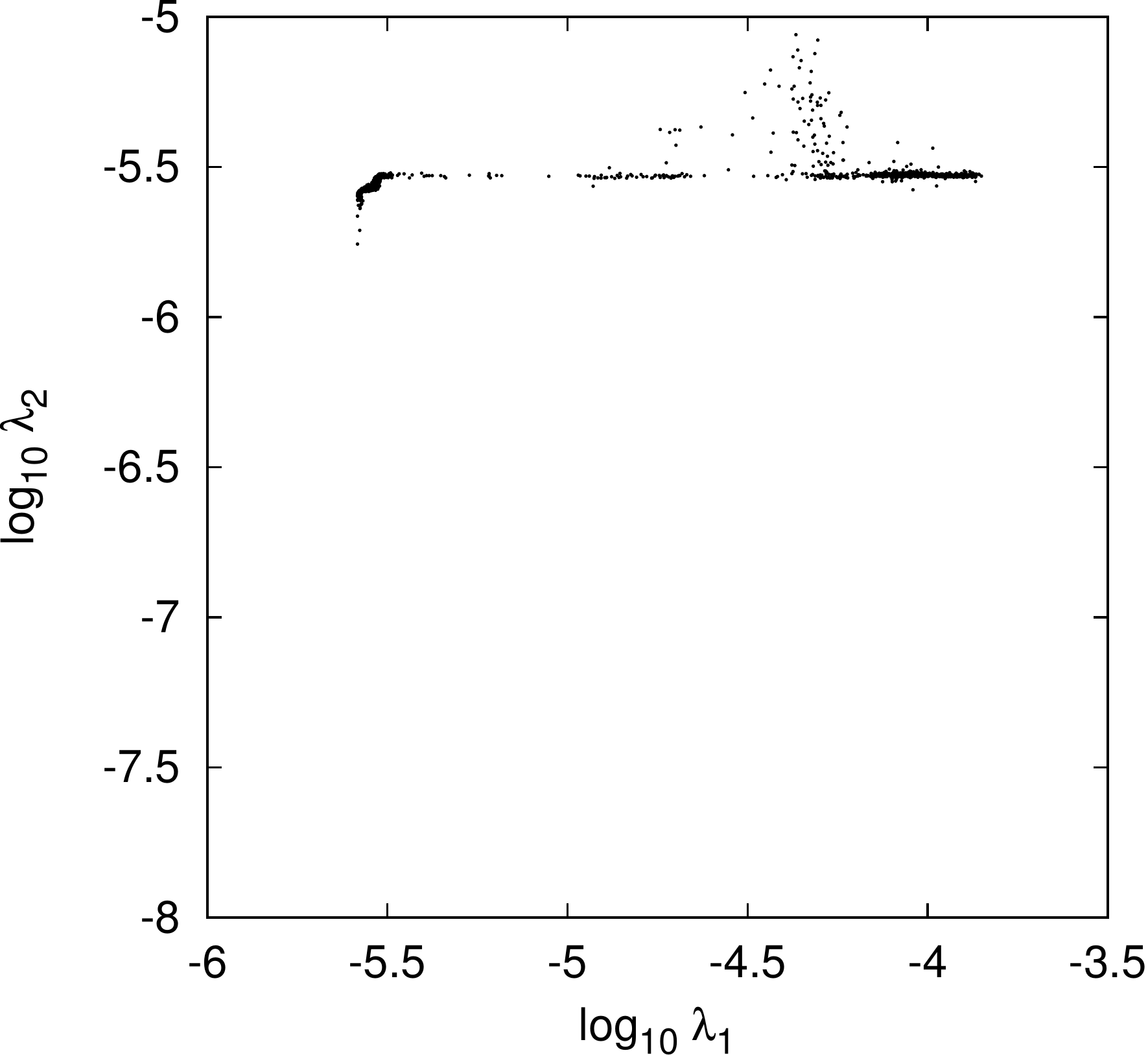}
\caption{Left: Logarithmic plot of the largest ($\lambda_1$) versus the second
largest ($\lambda_2$) LCEs
for a perturbed cubic oscilator (see text for details) using the standard ordering of
the deviation vectors. Right: the same but using the proposed ordering of
the deviation vectors.}
\label{ornotord}
\end{figure}

Besides, using the order of the initial vectors that we propose has the additional and more practically useful
advantage that it yields better defined values of the computed LCEs. As an example, we computed
the LCEs of orbits in the perturbed cubic force model used by
\citet{M17} with $-0.166843 < e_1 < -0.166507$,
$-0.324941 < e_2 < -0.323994$ and grid spacings $\Delta e_1=2^{-17}$ and
$\Delta e_2=2^{-16}$ (see \citet{M17} for details). The integrations were done
for $5\times 10^6$ time units in all the cases.
Figure~\ref{ornotord} presents the resulting $\lambda_1$ versus $\lambda_2$ with the usual ordering (left panel), and 
the same with our ordering (right panel). Clearly, the dispersion of the $\lambda_2$ values is smaller and their value better
defined with our ordering of the initial vectors.

\section{Conclusion}

The results obtained for the LCEs of a Hamiltonian autonomous system with the \citet{BGGS80b} method
depend on the order used for the initial deviation vectors. To obtain the expected opposite pairs of LCEs,
the deviation vectors should be ordered by pairs, the first, second, etc. ones having to be conjugate with
the last, the penultimate, etc., respectively. The results obtained with such an ordering have the
additional advantage of resulting in better defined values of the computed LCEs.

\section{Acknowledgements}

The authors are very grateful to Juani Rodr\'iguez for his technical support and they acknowledge financial support from the Universidad Nacional de La Plata, Argentina [Proyecto 11/G168] and from the Consejo Nacional de Investigaciones Cient\'ificas y T\'ecnicas, Argentina [PIP 2169].

\bibliography{biblio}

\begin{thebibliography}
\expandafter\ifx\csname natexlab\endcsname\relax\def\natexlab#1{#1}\fi
\expandafter\ifx\csname href\endcsname\relax
  \def\href#1#2{}\fi
\expandafter\ifx\csname urllinklabel\endcsname\relax
  \def\urllinklabel{[LINK]}\fi
\expandafter\ifx\csname adsurllinklabel\endcsname\relax
  \def\adsurllinklabel{[ADS]}\fi

\bibitem[{{Benettin} {et~al.}(1980{\natexlab{a}}){Benettin}, {Galgani},
  {Giorgilli}, \& {Strelcyn}}]{BGGS80a}
{Benettin}, G., {Galgani}, L., {Giorgilli}, A., \& {Strelcyn}, J.-M.
  1980{\natexlab{a}}, Meccanica, 15, 9


\bibitem[{{Benettin} {et~al.}(1980{\natexlab{b}}){Benettin}, {Galgani},
  {Giorgilli}, \& {Strelcyn}}]{BGGS80b}
---. 1980{\natexlab{b}}, Meccanica, 15, 21


\bibitem[{{Benettin} {et~al.}(1976){Benettin}, {Galgani}, \&
  {Strelcyn}}]{BGS76}
{Benettin}, G., {Galgani}, L., \& {Strelcyn}, J.-M. 1976, \pra, 14, 2338


\bibitem[{{Binney}(1982)}]{B82}
{Binney}, J. 1982, \mnras, 201, 1


\bibitem[{{Binney} \& {Spergel}(1982)}]{BS82}
{Binney}, J. \& {Spergel}, D. 1982, \apj, 252, 308


\bibitem[{{Carpintero} \& {Aguilar}(1998)}]{CA98}
{Carpintero}, D.~D. \& {Aguilar}, L.~A. 1998, \mnras, 298, 1


\bibitem[{{Carpintero} {et~al.}(2014){Carpintero}, {Maffione}, \&
  {Darriba}}]{CMD14}
{Carpintero}, D.~D., {Maffione}, N., \& {Darriba}, L. 2014, Astronomy and
  Computing, 5, 19


\bibitem[{{Carpintero} \& {Muzzio}(2016)}]{CM16}
{Carpintero}, D.~D. \& {Muzzio}, J.~C. 2016, \mnras, 459, 1082


\bibitem[{{Cincotta} {et~al.}(2003){Cincotta}, {Giordano}, \&
  {Sim{\'o}}}]{CGS03}
{Cincotta}, P.~M., {Giordano}, C.~M., \& {Sim{\'o}}, C. 2003, Physica D
  Nonlinear Phenomena, 182, 151


\bibitem[{{Cincotta} \& {Sim{\'o}}(2000)}]{CS00}
{Cincotta}, P.~M. \& {Sim{\'o}}, C. 2000, \aaps, 147, 205


\bibitem[{{Contopoulos}(2002)}]{C02}
{Contopoulos}, G. 2002, {Order and chaos in dynamical astronomy}


\bibitem[{{Contopoulos} \& {Voglis}(1996)}]{CV96}
{Contopoulos}, G. \& {Voglis}, N. 1996, Celest. Mech. Dynam. Astron., 64, 1


\bibitem[{{Darriba} {et~al.}(2012{\natexlab{a}}){Darriba}, {Maffione},
  {Cincotta}, \& {Giordano}}]{DMCG12p}
{Darriba}, L.~A., {Maffione}, N.~P., {Cincotta}, P.~M., \& {Giordano}, C.~M.
  2012{\natexlab{a}}, in 3rd La Plata International School on Astronomy and
  Geophysics: Chaos, diffusion and non-integrability in Hamiltonian systems -
  Application to astronomy, ed. C.~E. P.~M.~{Cincotta}, C. M.~{Giordano}
  (Universidad Nacional de La Plata - Asociaci\'on Argentina de Astronom\'\i
  a), 345--366


\bibitem[{{Darriba} {et~al.}(2012{\natexlab{b}}){Darriba}, {Maffione},
  {Cincotta}, \& {Giordano}}]{DMCG12}
{Darriba}, L.~A., {Maffione}, N.~P., {Cincotta}, P.~M., \& {Giordano}, C.~M.
  2012{\natexlab{b}}, International Journal of Bifurcation and Chaos, 22,
  1230033


\bibitem[{{Fouchard} {et~al.}(2002){Fouchard}, {Lega}, {Froeschl{\'e}}, \&
  {Froeschl{\'e}}}]{FLFF02}
{Fouchard}, M., {Lega}, E., {Froeschl{\'e}}, C., \& {Froeschl{\'e}}, C. 2002,
  Celest. Mech. Dynam. Astron., 83, 205


\bibitem[{{Froeschl{\'e}} {et~al.}(1997){Froeschl{\'e}}, {Gonczi}, \&
  {Lega}}]{FGL97}
{Froeschl{\'e}}, C., {Gonczi}, R., \& {Lega}, E. 1997, \planss, 45, 881


\bibitem[{Jackson(1989)}]{J89}
Jackson, E.~A. 1989, Perspectives of Nonlinear Dynamics, Vol.~1 (Cambridge
  University Press)


\bibitem[{{Laskar}(1990)}]{L90}
{Laskar}, J. 1990, \icarus, 88, 266


\bibitem[{{Lega} \& {Froeschl{\'e}}(2001)}]{LF01}
{Lega}, E. \& {Froeschl{\'e}}, C. 2001, Celest. Mech. Dynam. Astron., 81, 129


\bibitem[{{Maffione} {et~al.}(2011){Maffione}, {Darriba}, {Cincotta}, \&
  {Giordano}}]{MDCG11}
{Maffione}, N.~P., {Darriba}, L.~A., {Cincotta}, P.~M., \& {Giordano}, C.~M.
  2011, Celest. Mech. Dynam. Astron., 111, 285


\bibitem[{{Maffione} {et~al.}(2013){Maffione}, {Darriba}, {Cincotta}, \&
  {Giordano}}]{MDCG13}
---. 2013, \mnras, 429, 2700


\bibitem[{{Merritt} \& {Fridman}(1996)}]{MF96}
{Merritt}, D. \& {Fridman}, T. 1996, \apj, 460, 136


\bibitem[{{Muzzio}(2017)}]{M17}
{Muzzio}, J.~C. 2017, \mnras, 471, 4099


\bibitem[{{Papaphilippou} \& {Laskar}(1998)}]{PL98}
{Papaphilippou}, Y. \& {Laskar}, J. 1998, \aap, 329, 451


\bibitem[{{S{\'a}ndor} {et~al.}(2000){S{\'a}ndor}, {{\'E}rdi}, \&
  {Efthymiopoulos}}]{SEE00}
{S{\'a}ndor}, Z., {{\'E}rdi}, B., \& {Efthymiopoulos}, C. 2000, Celest. Mech.
  Dynam. Astron., 78, 113


\bibitem[{{S{\'a}ndor} {et~al.}(2004){S{\'a}ndor}, {{\'E}rdi}, {Sz{\'e}ll}, \&
  {Funk}}]{SESF04}
{S{\'a}ndor}, Z., {{\'E}rdi}, B., {Sz{\'e}ll}, A., \& {Funk}, B. 2004, Celest.
  Mech. Dynam. Astron., 90, 127


\bibitem[{{Skokos}(2001)}]{S01}
{Skokos}, C. 2001, Journal of Physics A: Mathematical and General, 34, 10029


\bibitem[{{Skokos} {et~al.}(2007){Skokos}, {Bountis}, \&
  {Antonopoulos}}]{SBA07}
{Skokos}, C., {Bountis}, T.~C., \& {Antonopoulos}, C. 2007, Physica D Nonlinear
  Phenomena, 231, 30


\bibitem[{Tabor(1989)}]{T89}
Tabor, M. 1989, Chaos and Integrability in Nonlinear Dynamics: An Introduction
  (Wiley)
 \href{https://books.google.com.ar/books?id=TkvvAAAAMAAJ}{\urllinklabel}

\bibitem[{{Udry} \& {Pfenniger}(1988)}]{UP88}
{Udry}, S. \& {Pfenniger}, D. 1988, \aap, 198, 135


\bibitem[{{{\v S}idlichovsk{\'y}} \& {Nesvorn{\'y}}(1996)}]{SN96}
{{\v S}idlichovsk{\'y}}, M. \& {Nesvorn{\'y}}, D. 1996, Celest. Mech. Dynam.
  Astron., 65, 137


\bibitem[{{Voglis} {et~al.}(1999){Voglis}, {Contopoulos}, \&
  {Efthymiopoulos}}]{VCE99}
{Voglis}, N., {Contopoulos}, G., \& {Efthymiopoulos}, C. 1999, Celest. Mech.
  Dynam. Astron., 73, 211


\bibitem[{{Voglis} \& {Contopoulos}(1994)}]{VC94}
{Voglis}, N. \& {Contopoulos}, G.~J. 1994, Journal of Physics A: Mathematical
  and General, 27, 4899


\end{thebibliography}

\end{document}